\documentclass[conference]{IEEEtran}
\usepackage{cite}

\ifCLASSINFOpdf
\usepackage[pdftex]{graphicx}
\else
\fi
\usepackage{amsmath}
\usepackage{amsfonts} 
\usepackage{amsthm}
\usepackage{amssymb}
\usepackage{amsthm} 
\usepackage{mathtools} 
\usepackage{textcomp} 
\usepackage{xcolor}
\usepackage{clrscode}

\usepackage{BOONDOX-ds}

\usepackage{bbm}

\newcommand{\pfun}{\mathop{\hbox{$\to$\kern-7pt\raise.9pt\hbox{\scalebox{1}[.55]{$|$}}\kern4pt} }}

\usepackage{array}

\begin{document}

\title{Analyzing 10 Petabit/s Network Data with Accelerated Associative (Token) Arrays
}

\author{\IEEEauthorblockN{
Jeremy Kepner, Hayden Jananthan, LaToya Anderson, William Arcand, David Bestor, William Bergeron, \\ 
Chansup Byun, Alex Bonn, Daniel Burrill, Vijay Gadepally, Michael Houle, Matthew Hubbell, Michael Jones, \\
Piotr Luszczek, Peter Michaleas, Lauren Milechin, Julie Mullen, Andrew Prout, Albert Reuther, \\
 Antonio Rosa, Charles Yee, Alex Pentland
\\
\IEEEauthorblockA{
MIT
}}}
\maketitle
\begin{abstract}
As networks expand and become an ever more critical infrastructure to modern society the need to analyze these networks with the highest regard for privacy is essential to ensure their proper function. Depending on the level of the network layer to be analyzed, sources and destinations can be any combination of physical, logical, or persona/agentic endpoints, which requires the ability to handle diverse data.  Invaluable to these analyses are mathematical tools that enable sophisticated mathematical algorithms to be expressed succinctly while achieving scalable vertical (within a compute node), horizontal (across compute nodes), and temporal (over different generations of hardware) performance. Associative (token) array mathematics and corresponding libraries is one approach that can meet these requirements.  Accelerating these libraries with GPUs enables the analysis of the largest networks. The MIT/IEEE/Amazon Anonymized Network Sensing Graph Challenge provides a venue for highlighting the applicability of accelerated associative arrays for these types of problems.  The D4M associative library has been implemented in a number of languages. This work benchmarks a prototype Matlab D4M GPU accelerated implementation of the Anonymized Network Sensing challenge across a wide range of CPU and GPU hardware.  Scalable performance is demonstrated within and across CPU cores, CPU nodes, and GPU nodes.  Horizontal scaling across multiple nodes was linear. Running on hundreds of GPU nodes simultaneously achieved a sustained processing rate sufficient to potentially analyze a 10 Petabit/s network.
\end{abstract}


%
\IEEEpeerreviewmaketitle

\section{Introduction}
\let\thefootnote\relax\footnotetext{Research was sponsored by the Department of the Air Force Artificial Intelligence Accelerator and was accomplished under Cooperative Agreement Number FA8750-19-2-1000. The views and conclusions contained in this document are those of the authors and should not be interpreted as representing the official policies, either expressed or implied, of the Department of the Air Force or the U.S. Government. The U.S. Government is authorized to reproduce and distribute reprints for Government purposes notwithstanding any copyright notation herein.
Use of this work is controlled by the human-to-human license listed in Exhibit 3 of https://doi.org/10.48550/arXiv.2306.09267
}

Many large-scale networking problems can only be solved with community access to very broad data sets with the highest regard for privacy and strong community buy-in \cite{kepner2021zero, pisharody2021realizing, pentland2022building}.  As networks expand and become an even more critical infrastructure for modern society, the need to analyze these networks is essential to ensure their proper function \cite{atkins2021improvised, atkins2021cooperation, demchak2021achieving, weed2022beyond, weed2023beyond, kepner2024normal, somin2025temporal}.  Mathematical tools that enable sophisticated mathematical algorithms to be expressed succinctly while achieving high scalability can play a key role in advancing the field. Associative (token) array mathematics \cite{kepnerjananthan} and corresponding libraries \cite{kepner2012dynamic, chen2016julia, jones2023deployment, quach2024integrating} accelerated with GPUs is one approach for assisting with the analysis of the largest networks.

The Anonymized Network Sensing Graph Challenge seeks to enable large, open, community-based approaches to protecting networks by providing a common venue for highlighting relevant innovations in network sensing and analysis \cite{jananthan2024anonymized, voloshchuk2025improving}.  This challenge is an ideal venue for exploring accelerated associative array mathematics for large-scale network analysis.

The presentation of the rest of the paper is as follows.  A more detailed description of the mathematical analysis performed within the Anonymized Network Sensing challenge is provided followed by a discussion of the mathematics of associative arrays.  Next, is a discussion of some of the key steps for adapting the specific associative array math library used in this work to take advantage of GPUs.  A key step for porting an application to GPUs is to port the underling basic functions; a short description of the functions for the corresponding associative array library is given.  Subsequently, the accelerated Anonymized Network Sensing challenge associative array code is presented.  This is followed by a description of the benchmarking hardware and the vertical (within node), horizontal (across nodes), and temporal (across time) scaling results.

\section{Anonymized Network Sensing Challenge}

The MIT/IEEE/Amazon GraphChallenge has fostered many community approaches for developing new solutions for analyzing graphs and sparse data derived from social media, sensor feeds, and scientific data to discover relationships between events as they unfold in the field \cite{pearce2017triangle, wolf2017fast, low2018linear, pearce2018ktruss, yasar2018fast, davis2019write, pandey2019hindex, pearce2019quadrillion, yasar2019linear, priest2020scaling, hidayetoglu2020atscale, ghosh2020tric, lin2020novel, xin2021fast, uppal2021faster, sun2022accelerating, xu2022towards, dun2023adaptive, wang2023smog, wanye2023integrated, lin2024mercury, qin2024towards, chen2025prism, dun2025towards}.  The anonymized network sensing Graph Challenge \cite{jananthan2024anonymized, voloshchuk2025improving} seeks to enable large, open, community-based approaches to protecting networks \cite{han2024extracting, lockton2025dbos, mandulak2025anonymized, milner2025interactive, samsi2025combining, wang2025sanst}. Many large-scale networking problems can only be solved with community access to very broad data sets with the highest regard for privacy and strong community buy-in. Such approaches often require community-based data sharing.  In the broader networking community (commercial, federal, and academia) anonymized source-to-destination traffic matrices with standard data sharing agreements have emerged as a data product that can meet many of these requirements.

\begin{figure}
\center{\includegraphics[width=1.0\columnwidth]{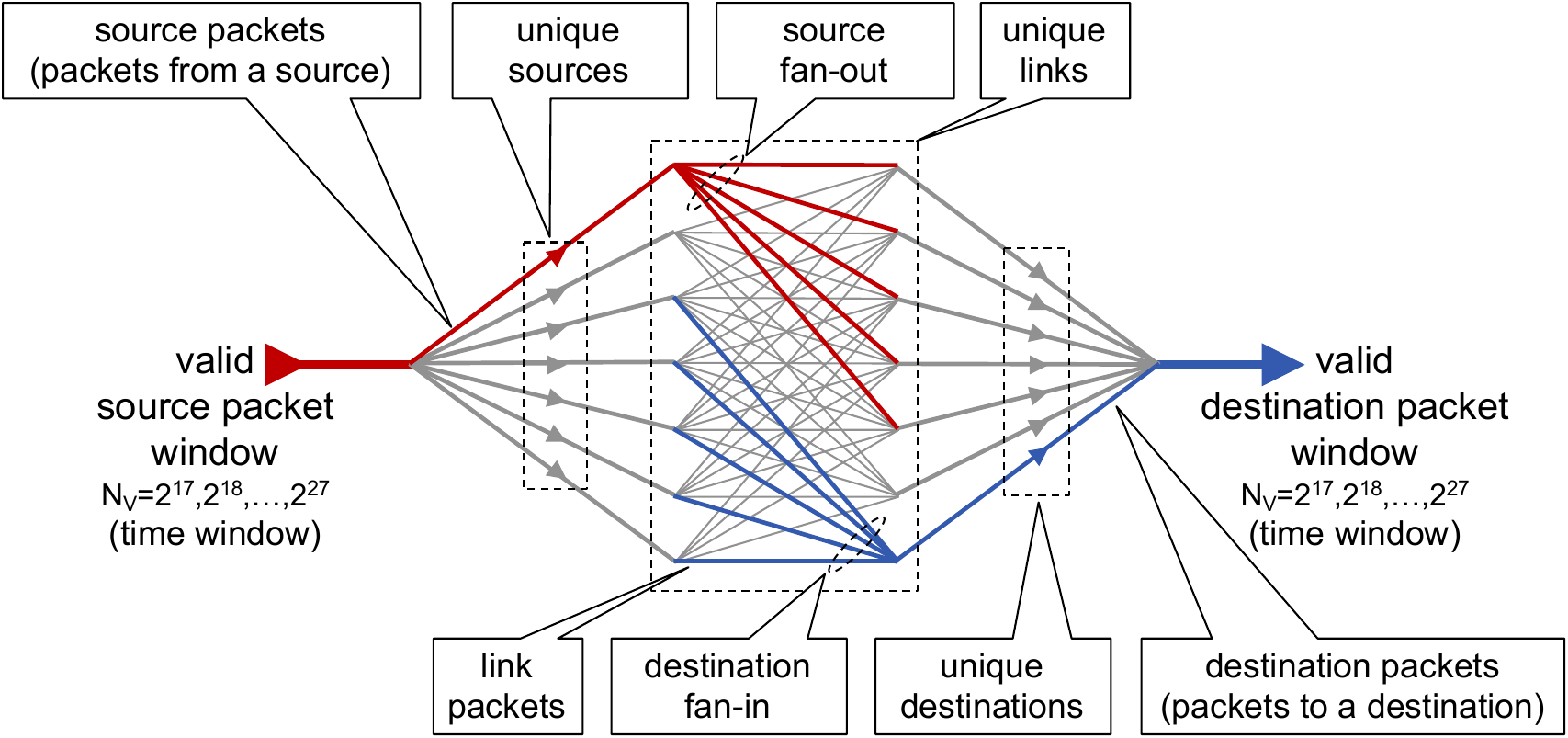}}
      	\caption{{\bf Anonymized Network Sensing Challenge Quantities.} Internet traffic streams of $N_V$ valid packets are divided into a variety of quantities for analysis by the anonymized network sensing challenge: source packets, source fan-out, unique source-destination pair packets (or links), destination fan-in, and destination packets.  Figure adapted from \cite{kepner19hypersparse}.}
      	\label{fig:NetworkDistribution}
\end{figure}
\begin{table}
\caption{Network Quantities from Traffic Matrices}
\vspace{-0.25cm}
Formulas for computing network quantities from a traffic matrix ${\bf A}_t$ at time $t$ in both summation and matrix notation. ${\bf 1}$ is a column vector of all 1's, $^{\sf T}$  is the transpose operation, and $|~|_0$ is the zero-norm that sets each nonzero value of its argument to 1\cite{karvanen2003measuring}.  These formulas are unaffected by matrix permutations and work on anonymized data.  \underline{\smash{Underlined}} quantities are those specified in the anonymized network sensing Graph Challenge. Table adapted from \cite{kepner2020multi}.
\begin{center}
\begin{tabular}{p{1.5in}p{0.9in}p{0.6in}}
\hline
{\bf Aggregate} & {\bf ~~~~Summation} & {\bf ~Matrix} \\
{\bf Property} & {\bf ~~~~~~Notation} & {\bf Notation} \\
\hline
\underline{\smash{Valid packets}} $N_V$ & $~\sum_i ~ \sum_j ~ {\bf A}_t(i,j)$ & $~{\bf 1}^{\sf T} {\bf A}_t {\bf 1}$ \\
\underline{\smash{Unique links}} & $~~\sum_i ~ \sum_j |{\bf A}_t(i,j)|_0$  & ${\bf 1}^{\sf T}|{\bf A}_t|_0 {\bf 1}$ \\
Link packets from $i$ to $j$ & $~~~~~~~~~~~~~~{\bf A}_t(i,j)$ & ~~~$~{\bf A}_t$ \\
\underline{\smash{Max link packets}} & $~~~~~\max_{ij}{\bf A}_t(i,j)$ & $\max({\bf A}_t)$ \\
\hline
\underline{\smash{Unique sources}} & $~\sum_i |\sum_j ~~ {\bf A}_t(i,j)|_0$  & ${\bf 1}^{\sf T}|{\bf A}_t {\bf 1}|_0$ \\
Packets from source $i$ & $~~~~~~~\sum_j ~ {\bf A}_t(i,j)$ & ~~$~~{\bf A}_t  {\bf 1}$ \\
\underline{\smash{Max source packets}}  & $ \max_i \sum_j ~ {\bf A}_t(i,j)$ & $\max({\bf A}_t {\bf 1})$ \\
Source fan-out from $i$ & $~~~~~~~~~~~\sum_j |{\bf A}_t(i,j)|_0$  & ~~~$|{\bf A}_t|_0 {\bf 1}$ \\
\underline{\smash{Max source fan-out}} & $ \max_i \sum_j |{\bf A}_t(i,j)|_0$  & $\max(|{\bf A}_t|_0 {\bf 1})$ \\
\hline
\underline{\smash{Unique destinations}} & $~\sum_j |\sum_i ~ {\bf A}_t(i,j)|_0$ & $|{\bf 1}^{\sf T} {\bf A}_t|_0 {\bf 1}$ \\
Destination packets to $j$ & $~~~~~~~\sum_i ~ {\bf A}_t(i,j)$ & ${\bf 1}^{\sf T}~{\bf A}_t$ \\
\underline{\smash{Max destination packets}} & $ \max_j \sum_i ~ {\bf A}_t(i,j)$ & $\max({\bf 1}^{\sf T}~{\bf A}_t)$ \\
Destination fan-in to $j$ & $~~~~~~~~~~~\sum_i |{\bf A}_t(i,j)|_0$ & ${\bf 1}^{\sf T}|{\bf A}_t|_0$ \\
\underline{\smash{Max destination fan-in}} & $ \max_j \sum_i |{\bf A}_t(i,j)|_0$ & $\max({\bf 1}^{\sf T}|{\bf A}_t|_0)$ \\
\hline
\end{tabular}
\end{center}
\label{tab:Aggregates}
\end{table}

A core concept in the Anonymized Network Sensing challenge is that traffic data can be viewed as a traffic matrix where each row is a source and each column is a destination. A primary benefit of constructing anonymized  traffic matrices is the efficient computation of a wide range of network quantities via matrix mathematics. Figure~\ref{fig:NetworkDistribution} illustrates essential quantities found in all streaming dynamic networks. These quantities are all computable from anonymized traffic matrices created from the source and destination addresses found in Internet packet headers \cite{soule2004identify, zhang2005estimating, mucha2010community, tune2013internet}. To reduce statistical fluctuations, the streaming data is partitioned so that for any chosen time window all data sets have the same number of valid packets \cite{kepner2020multi}.  At a given time $t$, $N_V$ consecutive valid packets are aggregated from the network traffic into a  matrix ${\bf A}_t$, where ${\bf A}_t(i,j)$ is the number of valid packets between the source $i$ and destination $j$. The sum of all the entries in ${\bf A}_t$ is equal to $N_V$
$$
    \sum_{i,j} {\bf A}_t(i,j) = N_V
$$
Constant packet, variable time samples simplify the statistical analysis of the heavy-tail distributions commonly found in network traffic quantities \cite{kepner19hypersparse, nair2020fundamentals, kepner2022new}.  All the network quantities depicted in Figure~\ref{fig:NetworkDistribution} can be readily computed from ${\bf A}_t$ using the formulas listed in Table~\ref{tab:Aggregates}.

\section{Associative (Token) Arrays}

Depending on the level of the network layer to be analyzed, sources and destinations can be any combination of physical, logical, or persona/agentic endpoints, such as, AI agents communicating via MCP (Model Context Protocol) \cite{south2025identity}.  These endpoints are often represented as strings.  Associative  arrays, generalize matrices and their operations to much broader data sets \cite{kepnerjananthan}.  More specifically, an array
$$
   {\bf A}(i,j) = v
$$
is a mapping from a pair of indices $i$ and $j$ to a value $v$ where
$$
    i \in I, ~~ j \in J,  ~~ v \in V
$$
For an $N{\times}N$ real-valued matrix ${\bf A} : \mathbb{R}^{N{\times}N}$
$$
    I = J = \{1,...,N\}, ~~ V = \mathbb{R}
$$
An associative array generalizes this concept by allowing $I, J, V$ to be any strict totally ordered set, which includes numbers and strings.  In AI large language model (LLM) terminology, strict totally ordered sets are simply tokens; often words or parts of words that are used as the labels of the rows and columns of the many matrices used insides LLMs.  In many respects, LLMs are simply collections of associative (token) arrays.

In the Anonymized Network Sensing Graph Challenge the endpoints are set by Internet Protocol version 4 (IPv4) using 32 bit unsigned numbers. An IPv4 source/destination can be displayed as an integer (67305985), hexadecimal value (01020304), binary value (00000001000000100000001100000100), padded dotted quad string (001.002.003.004), or unpadded dotted quad string (1.2.3.4). Associative arrays encompass all of these and allow the same mathematics to be performed regardless of the underlying representation.  The Dynamic Distributed Dimensional Data Model (D4M) mathematical library (see d4m.mit.edu) implements associative arrays in a number of high-level programming languages \cite{kepner2012dynamic, chen2016julia, jones2023deployment, quach2024integrating}.

\section{Accelerated Arrays}

Porting D4M to a GPU enables all of these representations to benefit from accelerated computing. The focus of this work is on the Matlab D4M implementation that leverages the extensive GPU programming system within Matlab.  A cornerstone of the Matlab GPU system are two functions {\tt gpuArray()} and {\tt gather()} that respectively copy a Matlab variable from a CPU to a CPU and from GPU to a CPU 
\vspace{0.1cm}

{\tt\small Agpu = gpuArray(Acpu)}

- Copies variable to GPU

 - Changes type to GPU variable

 {\tt\small Acpu = gather(Agpu)}

 - Copies from GPU

 - Changes type to CPU variable

\vspace{0.1cm}

\noindent In addition, the type of the variable is also changed from a CPU variable to GPU variable and from GPU variable to a CPU variable, respectively.  Thus, a Matlab GPU program can often appear identical to a Matlab CPU program as type information determines whether any given function is run on a CPU or a GPU. Thus, a Matlab accelerated GPU program can often appear identical to a normal Matlab CPU program as type information determines whether any given function is run on a CPU or a GPU.

D4M employs object type and function overloading in a similar manner.  The core D4M associative array data structure can be converted to a GPU object via corresponding function overloading 

\vspace{0.1cm}

{\tt\small AgpuD4M = gpuArray(AcpuD4M)}

~~{\tt\small AgpuD4M.row = gpuArray(uint16(AcpuD4M.row))}

~~{\tt\small AgpuD4M.col = gpuArray(uint16(AcpuD4M.col))}

~~{\tt\small AgpuD4M.val = gpuArray(uint16(AcpuD4M.val))}

~~{\tt\small AgpuD4M.A = gpuArray(AcpuD4M.A)}

\vspace{0.1cm}

{\tt\small AcpuD4M = gather(AgpuD4M)}

~~{\tt\small AcpuD4M.row = char(gather(AgpuD4M.row))}

~~{\tt\small AcpuD4M.col = char(gather(AgpuD4M.col))}

~~{\tt\small AcpuD4M.val = char(gather(AgpuD4M.val))}

~~{\tt\small AcpuD4M.A = gather(AgpuD4M.A)}

\vspace{0.1cm}

\noindent Subsequent D4M operations on these fields will then be performed on the GPU.  The D4M data structure consists of three character array fields ({\tt\small .row}, {\tt\small .col}, {\tt\small .val}) and a sparse matrix ({\tt\small .A}) holding an index to the corresponding entry in {\tt\small .val}) [Note: if {\tt\small .val} is empty, then the associative array values are numeric and are as given in {\tt\small .A}].  An important choice in the original D4M design was to store character array fields as flat vectors using the convention that the last character in the vector was the string separator.  This enabled operations on these fields to be performed as vector operations that are performant on both CPUs and GPUs.  At the time of writing, the Matlab character type was not directly supported on the GPU, so it is converted to a GPU supported unsigned 16-bit integer.   Porting of D4M functions to operate on GPU associative arrays is mostly an exercise in rewriting code to use performant functions that are supported on both the CPU and GPU, which allows the code to be independent of where it is run.  
Perhaps the most common trick used to achieve this end is rewriting array index operations, which have limited GPU support on sparse matrices, to employ sparse matrix multiply of corresponding diagonal matrices
$$
 {\bf A}({\bf i},{\bf j}) = {\bf I}_{\rm diag} ~ {\bf A} ~ {\bf J}_{\rm diag}
$$
where ${\bf I}_{\rm diag}$ has 1's along the diagonal corresponding to the rows to be selected (i.e., ${\bf I}_{\rm diag}({\bf i}(i),{\bf i}(i)) = 1$), and ${\bf J}_{\rm diag}$ has 1's along the diagonal corresponding to the columns to be selected (i.e., ${\bf J}_{\rm diag}({\bf j}(j),{\bf i}(j)) = 1$). 
The above approach generally results in improved performance on both CPUs and GPUs, while often allowing very complicated index remapping operations to be combined into a single step.  Sparse matrix libraries such as the GraphBLAS use this method to implement matrix indexing as it a leverages the already heavily optimized matrix multiplication function.  D4M also uses matrix multiplication to implement indexing, thus D4M indexing function itself required little adjustment to run on a GPU.  Generally speaking, the more vectorized the code is, the more it will run well on both CPUs and GPUs without modification.

It is particularly worth noting the impressive Matlab profiling environment that works identically for both CPU and GPU code.  The profiler clearly highlights the most time consuming parts of the code.  It is then simple to insert a breakpoint before the corresponding code and perform live \emph{in situ} performance tuning to optimize the code. The profiler makes Matlab a highly effective environment for discovering optimal GPU algorithms for subsequent translation into other languages.

\begin{table}
\caption{Basic Associative Array Operations}
\begin{center}
\begin{tabular}{ll}
\hline
Construction & {\tt\small A = Assoc(row,col,val)} \\
Addition & {\tt\small  C = A + B} \\
Subtraction & {\tt\small C = A - B} \\
Array Multiplication & {\tt\small  C = A * B} \\
Elementwise Multiplication & {\tt\small C = A .* B} \\
\hline
\end{tabular}
\end{center}
\label{tab:BasicOperations}
\end{table}

\section{Basic Functions}

Associative array libraries rely on a few basic functions for implementing algorithms (see Table~\ref{tab:BasicOperations}).  Initial GPU porting and benchmarking starts with these functions.  An important aspect of benchmarking these functions is the data used to randomly populate the appropriate sparse associative arrays.  Two important parameters are the sizes of $N{\times}N$ associative arrays and the number of non-empty entries $M$.  A common way to connect these parameters is via the formula $M = k N$, where $k \sim 10$.  Finally, the associative array indices are stored as their decimal character representation padded to the length of $N$.  For example, for $N=2^{10}$ the indices would be drawn from set of strings $\{0001,...,1024\}$.

\section{Benchmark Code}

The Anonymized Network Sensing Graph Challenge consists of six steps
\begin{enumerate}
\item Read/stream each network packet capture (PCAP) file.
\item Extract the source IP and destination IP addresses from the packet headers and buffer $N_V$ valid packets.
\item Anonymize the source IP and destination IP.
\item Construct sequential traffic matrices from $N_V$ valid packets.
\item Save the traffic matrices to files.
\item Read in the traffic matrix files, sum the traffic matrices associated with a PCAP file into one large traffic matrix, and perform the analysis highlighted in Table~\ref{tab:Aggregates}.
\end{enumerate}
Furthermore, ``Graph Challenge participants are free to select (with accompanying explanation) the Graph Challenge elements  that are appropriate for highlighting their innovations.''\cite{jananthan2024anonymized}  Along these lines, this work uses the Matlab reference code available on the GraphChallenge.org website and adapts it to use associative arrays.  The benchmarking focus of the work is on the analysis part of Step (6).  Thus, the code begins by reading in GraphBLAS versions of the random traffic matrices that are available from the GraphChallenge.org website.  The matrices are converted to associative arrays and then the analysis are executed and timed.

The GraphBLAS traffic matrix data was converted to associative arrays by extracting the row, column, and value triples. For the CPU, the rows/columns were converted to padded dotted quad representations used to construct corresponding associative arrays.  For the GPU, since strings are already being converted to 16-bit unsigned integers an additional optimization was be performed by converting each byte of the row/column IPv4 address to 16-bit unsigned integer (offset by 1 to avoid the side effects of using 0 when integrating with sparse libraries). This results in the GPU indices being $\sim$1/3 the size of the CPU indices.

The analysis code operations corresponding to Table~\ref{tab:Aggregates} were implemented as follows using the same code on CPUs and GPUs

\vspace{0.1cm}

{\tt\small AA = Adj(A)}

{\tt\small v = nonzeros(AA)}

{\tt\small \underline{Npackets} = sum(v)}

{\tt\small \underline{Nlinks} = length(v)}                                         

{\tt\small \underline{Nsrc} = size(A,1)}                                    

{\tt\small \underline{Ndest} = size(A,2)}                                

{\tt\small \underline{MaxPackets} = max(v)}                               

{\tt\small colVec = ones(Ndest,1,localType)}

{\tt\small \underline{MaxSrcPackets} = max(full(AA*colVec))}

{\tt\small \underline{MaxFanOut} = max(full(sign(AA)*colVec))}             

{\tt\small rowVec = ones(1,Nsrc,localType)}

{\tt\small \underline{MaxDestPackets} = max(full(rowVec*AA))}             

{\tt\small \underline{MaxFanIn} = max(full(rowVec*sign(AA)))}

\vspace{0.1cm}

\noindent The above code is performant on both CPUs and GPUs.  The potential GPU nature of the code is only manifest in the use of the {\tt\small localType} variable that instructs the {\tt\small ones()} constructor where to create data.  The code simultaneously uses the associative array {\tt\small A}, sparse adjacency matrix {\tt\small AA}, and value list {\tt\small v} representations of the data to provide a variety of options for selecting operations that perform well on both CPUs and the GPUs.  For example, summing and maxing {\tt\small v} as the fastest way to compute {\tt\small\underline{Npackets}} and {\tt\small\underline{Nlinks}}.  Likewise, multiplying {\tt\small AA} by a dense row/column vectors to compute {\tt\small\underline{MaxSrcPackets}}, {\tt\small\underline{MaxFanOut}}, {\tt\small\underline{MaxDestPackets}}, and {\tt\small\underline{MaxFanIn}}.

\section{Benchmarking}

Our team has developed a high-productivity scalable platform---the MIT SuperCloud---for providing scientists and engineers the tools they need to analyze large-scale dynamic data \cite{kepner2012dynamic, gadepally2018hyperscaling, 8547629}.  The MIT SuperCloud provides interactive analysis capabilities  accessible from high level programming environments that scale to thousands of processing nodes.  MIT SuperCloud maintains a diverse set of hardware running an identical software stack that allows direct comparison of hardware from different eras.

\begin{table}
\caption{Computer Hardware Specifications}
\vspace{-0.25cm}
MIT SuperCloud maintains a diverse set of hardware running an identical modern software stack providing a unique platform for comparing performance over different eras.
\begin{center}
\includegraphics[width=\columnwidth]{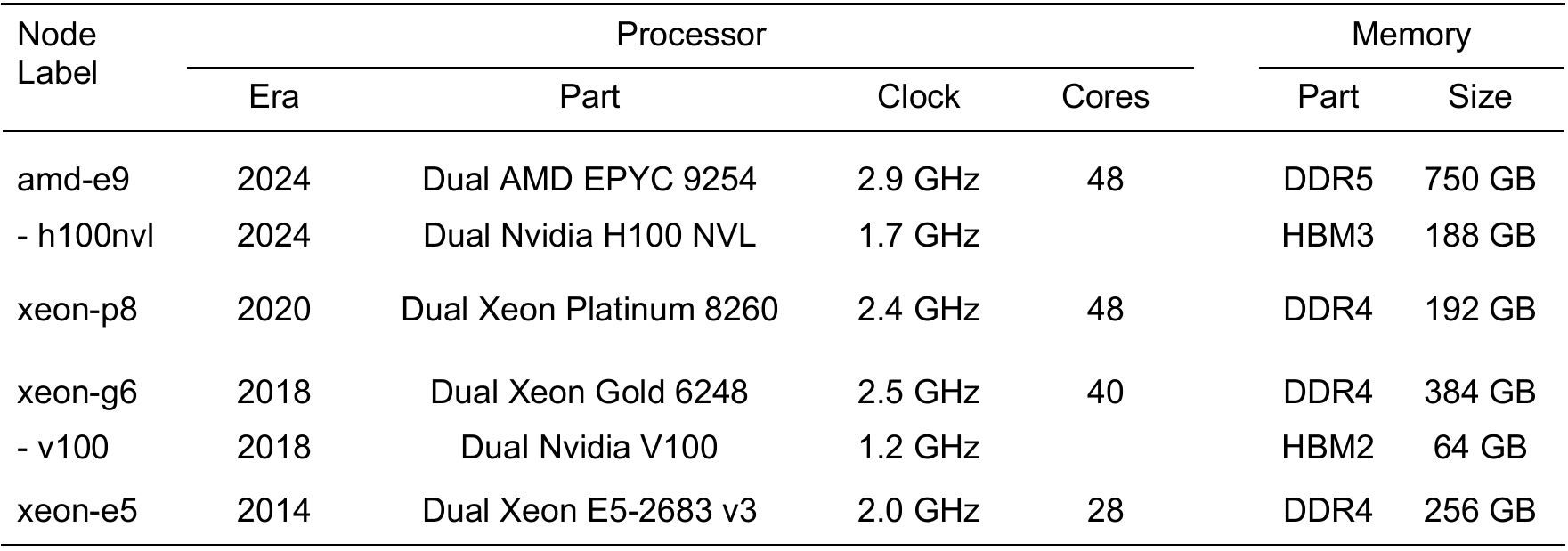}
\end{center}
\label{tab:HardwareTable}
\end{table}

\begin{table}
\caption{Single Node Basic Functions Parameters}
\vspace{-0.25cm}
Number of processes ($N_P$), number of cores/GPUs per process ($N_{CPP}$), array dimension $N$, and average number of non-empty entries per row/column ($k$) used for single node benchmarking of the associative array basic functions.  These benchmark parameters were chosen to balance consistency with the memory capacity of the different hardware. 
\begin{center}
\includegraphics[width=\columnwidth]{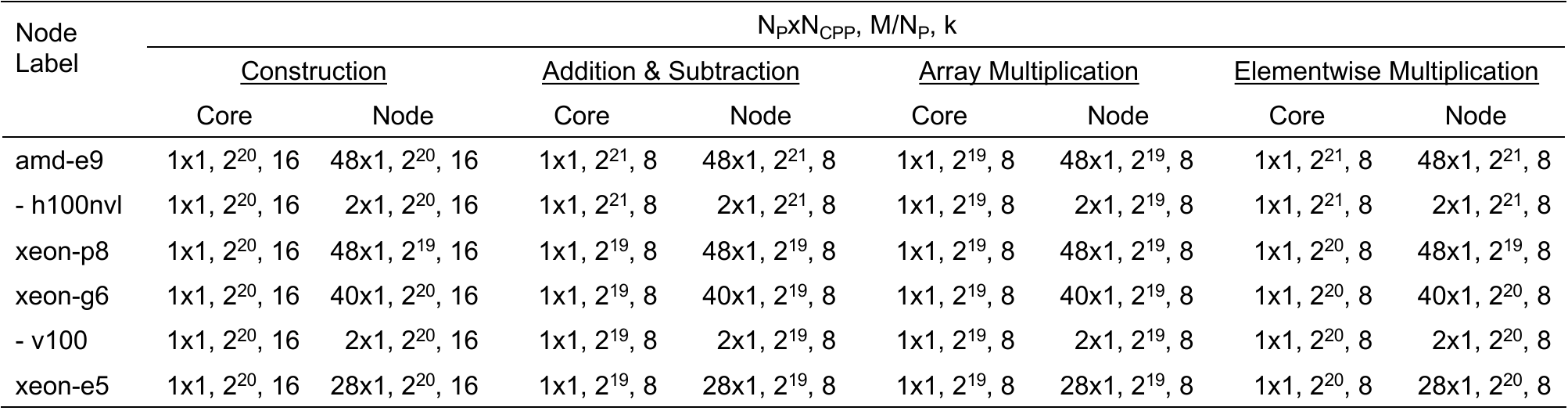}
\end{center}
\label{tab:BasicFunctions-parameters}
\end{table}

\begin{table}
\caption{Single Node Anonymized Network Sensing Parameters}
\vspace{-0.25cm}
Number of processes ($N_P$), number of cores/GPUs per process ($N_{CPP}$) and total packets per process $N_V/N_P$ used for single node benchmarking.  Anonymized network sensing benchmark parameters were chosen to balance consistency with the memory capacity of the different hardware.  For multiple nodes the parameters highlighted in {\bf bold} were used.
\begin{center}
\includegraphics[width=\columnwidth]{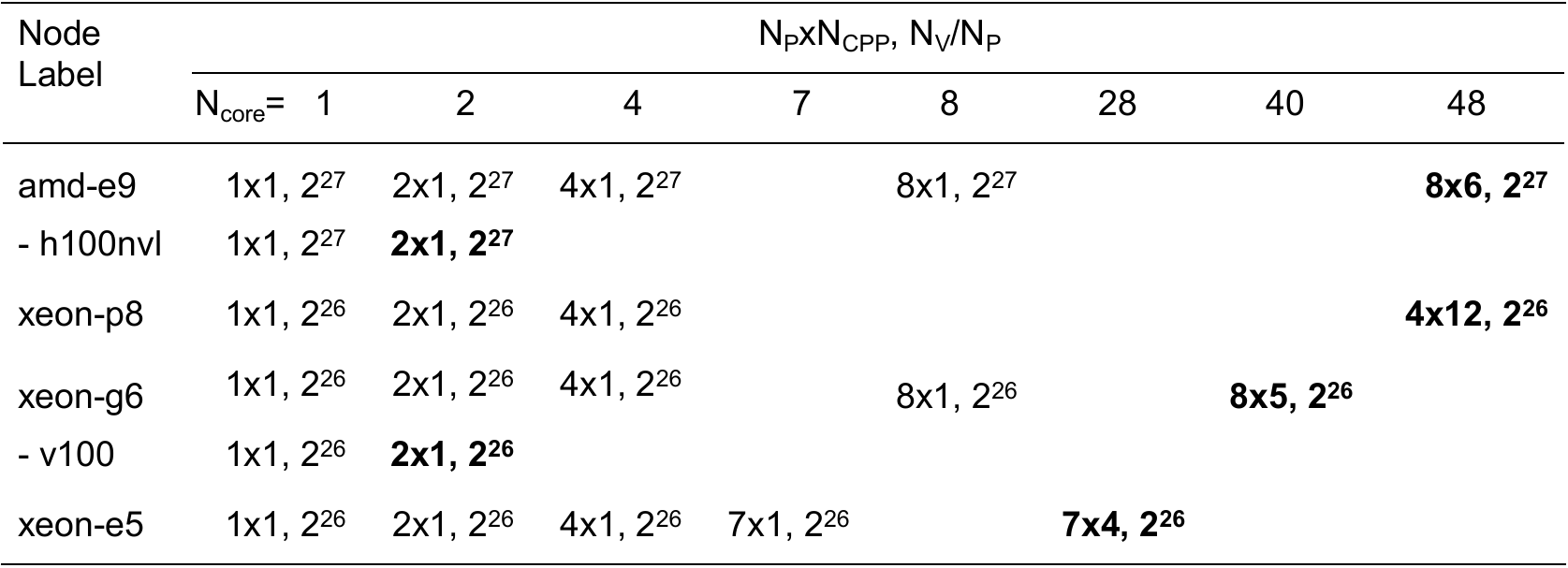}
\end{center}
\label{tab:AnonNetSense-parameters}
\end{table}

\begin{figure*}[]
\centering
\includegraphics[width=0.9\columnwidth]{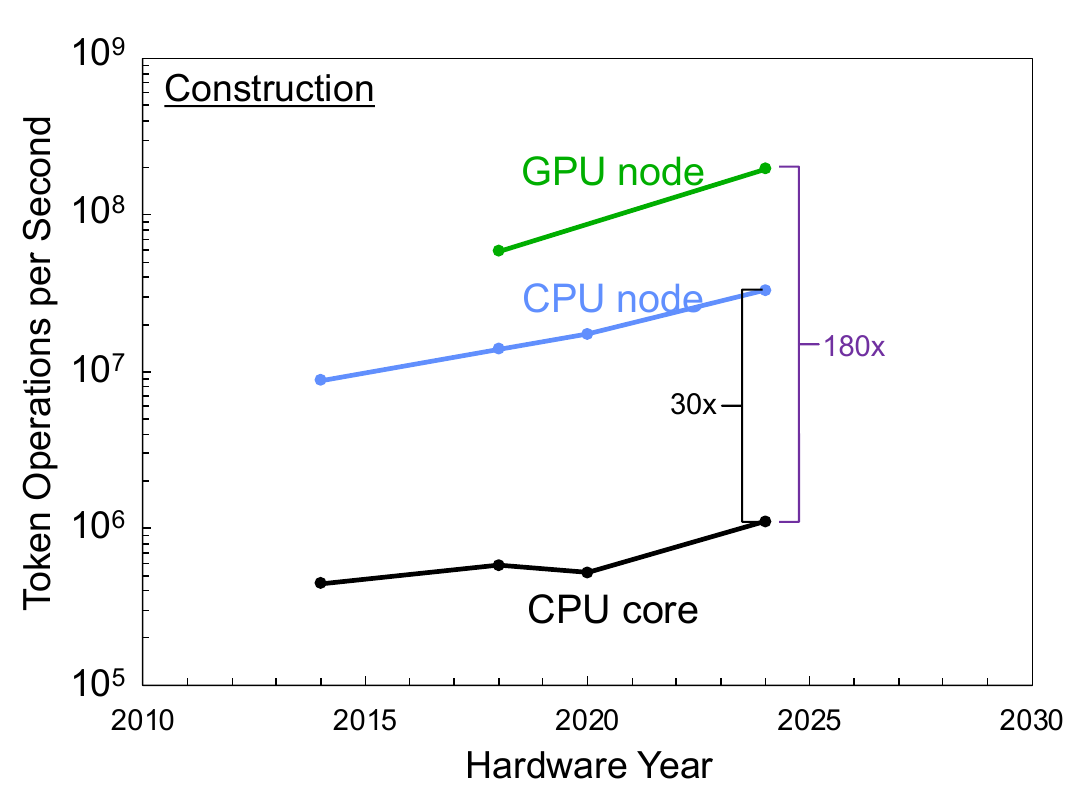}
\\
\includegraphics[width=0.9\columnwidth]{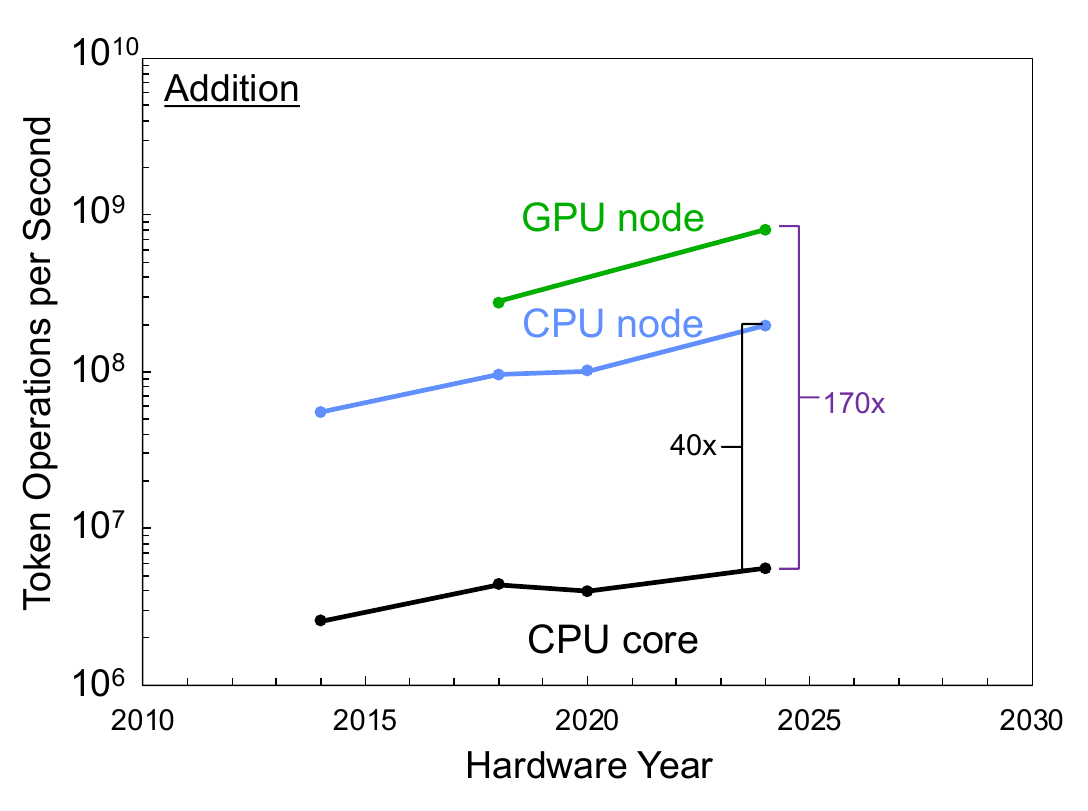}
\includegraphics[width=0.9\columnwidth]{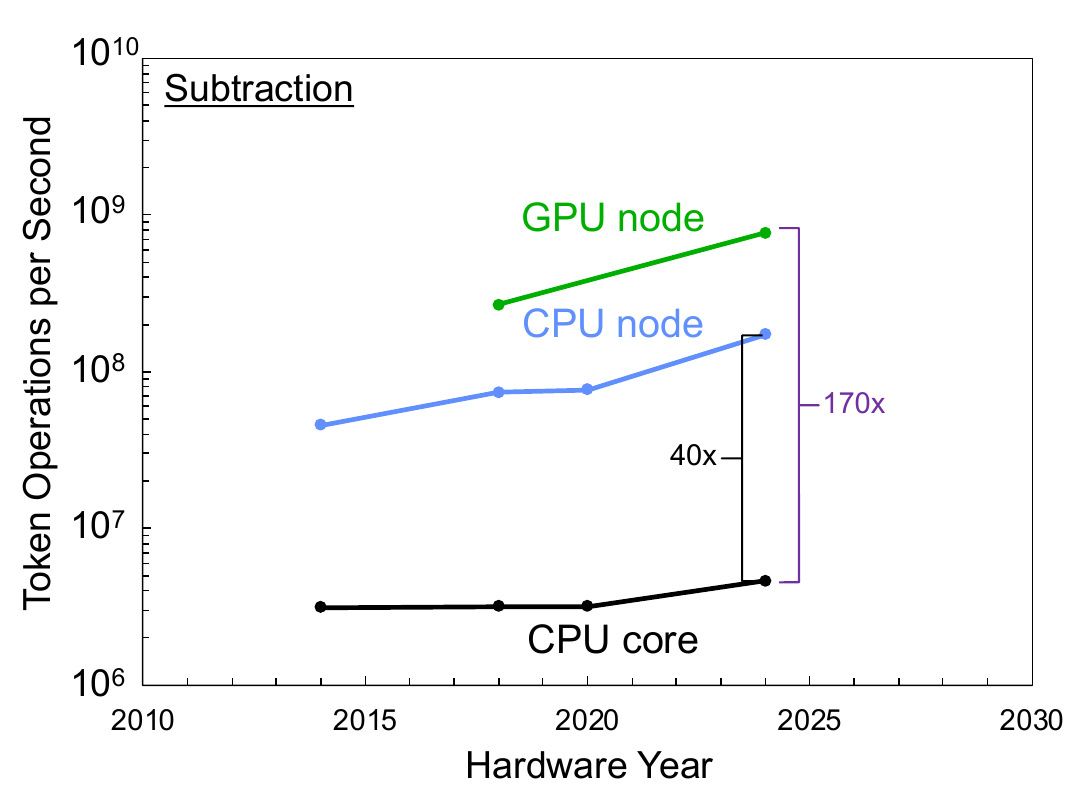}
\includegraphics[width=0.9\columnwidth]{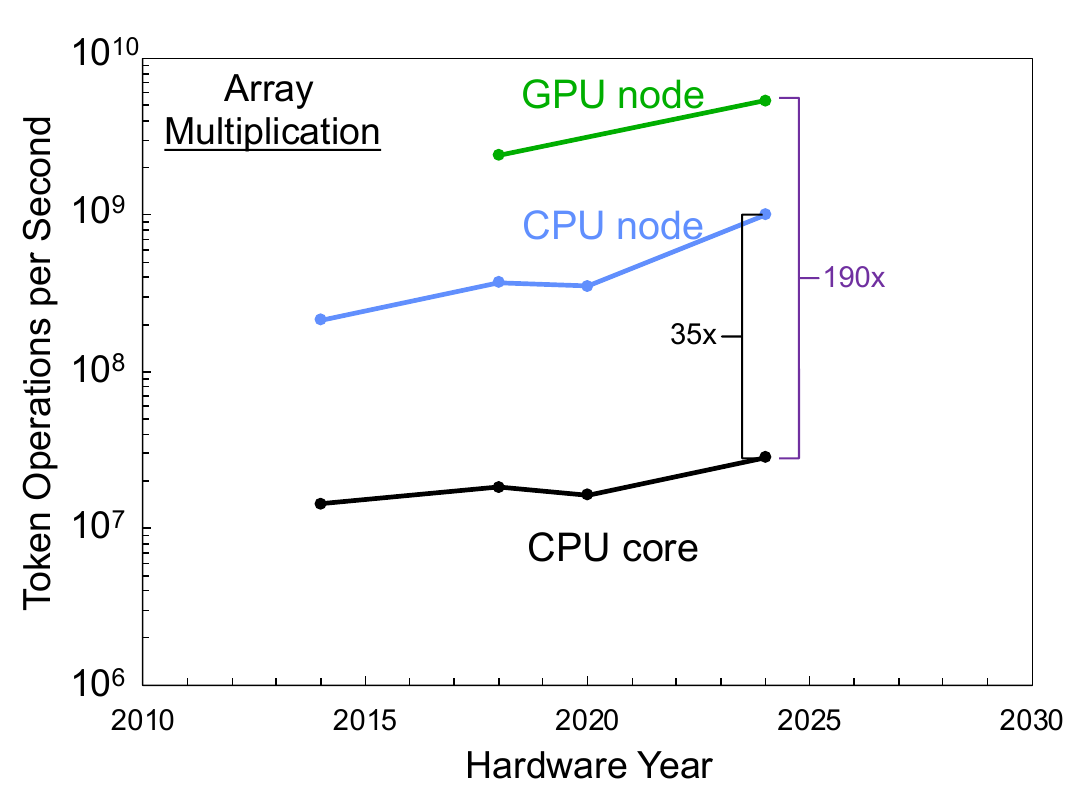}
\includegraphics[width=0.9\columnwidth]{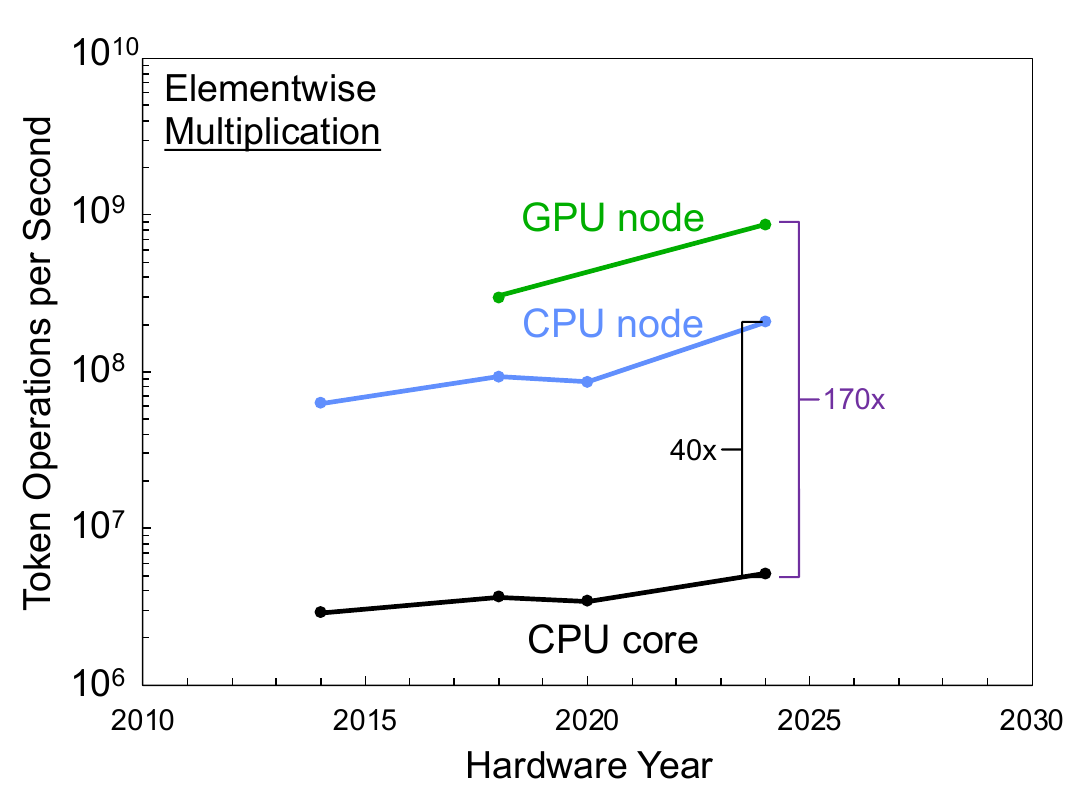}
\caption{{\bf Associative (Token) Array Basic Functions}.  Performance is measured for the different hardware configurations (see Table~\ref{tab:HardwareTable}).  All plots show consistent excellent vertical scaling within a node (approximately linear in number of CPU cores or GPUs) and temporal scaling over multiple eras of hardware.}
\label{fig:BasicFunctions}
\end{figure*}

\begin{figure*}[]
\centering
\includegraphics[width=0.86\columnwidth]{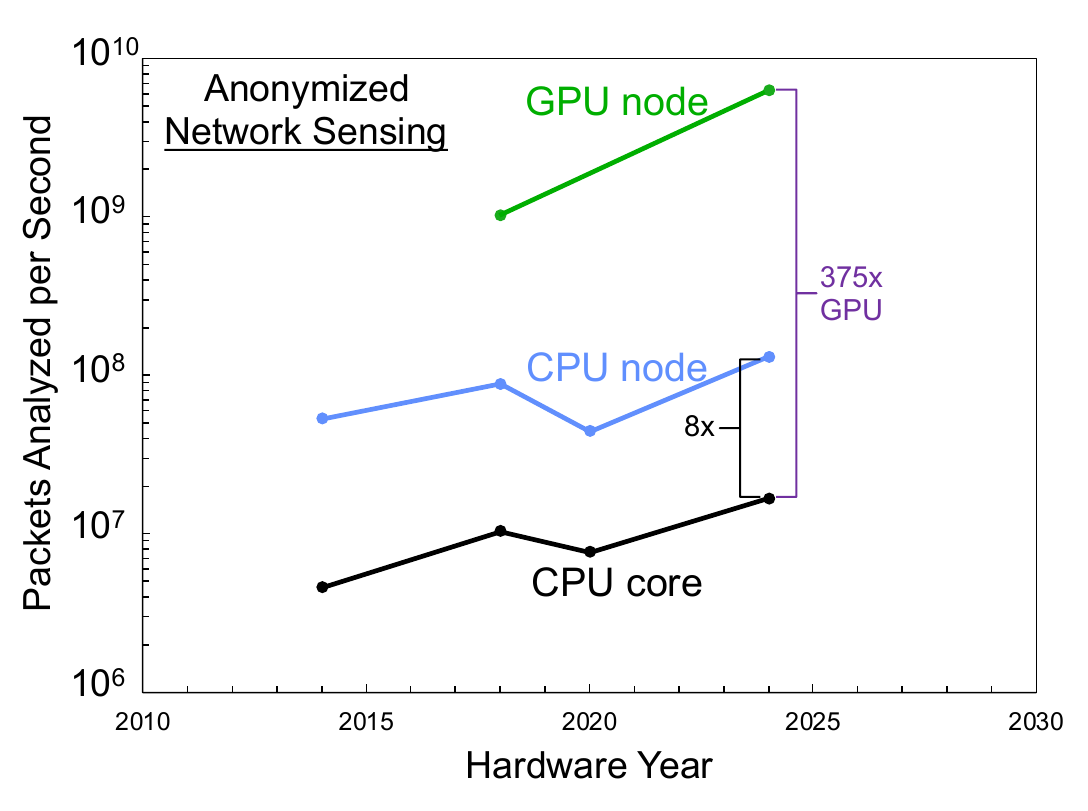}
\includegraphics[width=0.94\columnwidth]{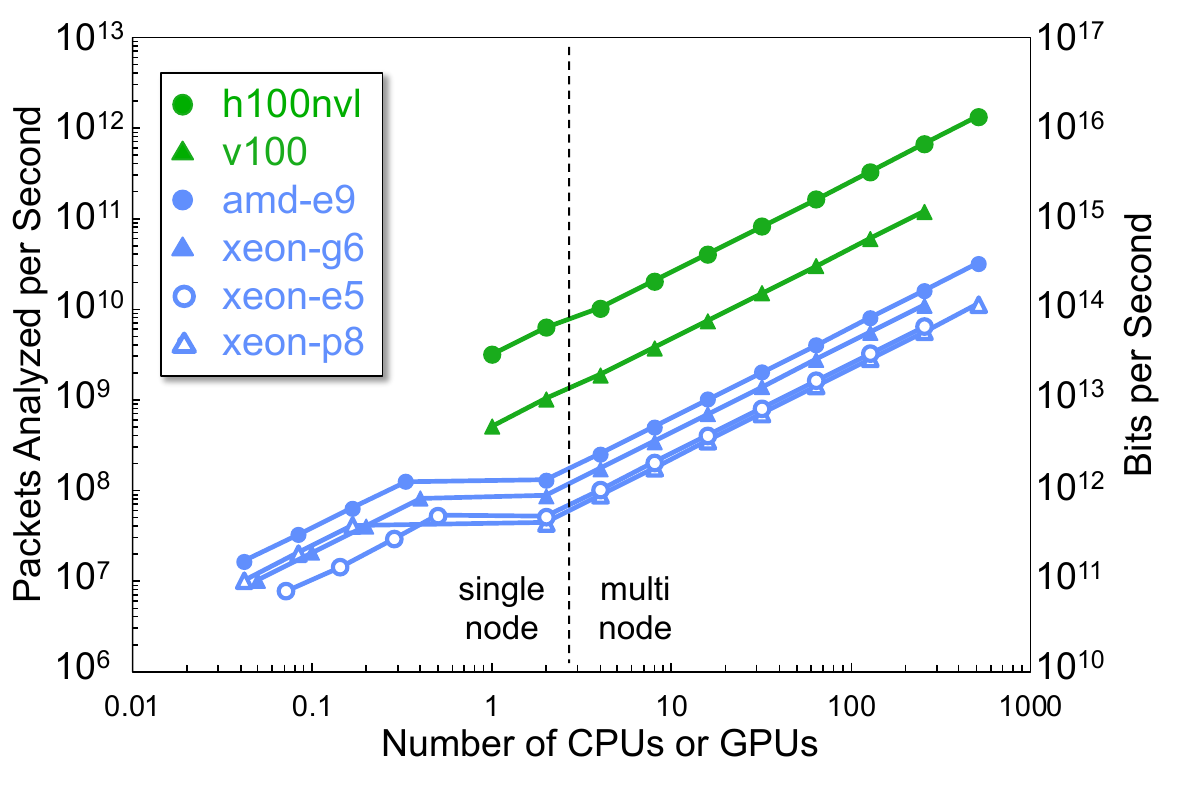}
\caption{{\bf Anonymized Network Sensing Challenge Temporal, Vertical, and Horizontal Scaling}. Performance is measured for the different hardware configurations (see Table~\ref{tab:HardwareTable}) using the random data sets available from GraphBLAS.org.  All plots show consistent excellent vertical scaling within a node (approximately linear in number of CPU cores or GPUs up to memory limits) and temporal scaling over multiple eras of hardware (left panel).  Likewise, the horizontal scaling is linear with the number of CPUs or GPUs (right panel).  Note: for CPUs a single socket has a value of 1 on the x-axis; using a single core on a 24 core CPU would be 1/24 = 0.041 of a CPU.
}
\label{fig:AnonNetSensing-scaling}
\end{figure*}

A typical benchmarking run can be launched in a few seconds using the MIT SuperCloud using the triples-mode hierarchical launching system \cite{8547629} enabling rapid interactive benchmarking.  The launch parameters are [$N_{node}$ $N_{ppn}$ $N_{tpn}$], corresponding to $N_{node}$ nodes, $N_{ppn}$ Matlab processes per node, and $N_{tpn}$ OpenMP threads per process.   The total number of processes is given by $N_P = N_{node} N_{ppn}$. On each node, each of the $N_{ppn}$ processes and their corresponding $N_{tpn}$ threads were pinned to adjacent cores to minimize interprocess contention and maximize cache locality \cite{byun2019optimizing}.  Within each Matlab process, OpenMP parallelism is used as provided by the underling math libraries.  The number hardware cores/GPUs per process is given by $N_{CPP}$.  $N_{CPP} = N_{tpn}$ on CPUs and $N_{CPP} = 1$ on GPUs. Triples mode makes it easy to explore horizontal scaling  across nodes, vertical scaling by examining combinations of processes and threads on a node, and temporal scaling by running on diverse hardware from different eras.

The computing hardware consists of many different types of nodes spanning over a decade (see Table~\ref{tab:HardwareTable}). The MIT SuperCloud maintains the same modern software stack across all nodes, which allows for direct comparison of hardware performance differences.

The parameters used to benchmark the associative array Basic Functions are shown in Table~\ref{tab:BasicFunctions-parameters}.  These benchmark parameters were chosen to balance consistency with the memory capacity of the different hardware.  In general, the associative arrays sizes $N$ were $10^{19}$, $10^{20}$, or $10^{21}$ and the average non-empty entries per row/column $k$ was 8 or 16. The memory intensive nature of these operations meant that the optimal whole node performance is achieved by running a number of processes equal to the number of cores or GPUs on the node.

The parameters used to benchmark the associative array implementation of the Anonymized Network Sensing challenge are shown in Table~\ref{tab:AnonNetSense-parameters}. These benchmark parameters were chosen to balance consistency with the memory capacity of the different hardware.  The Anonymized Network Sensing challenge data sets are large and smaller memory nodes could only run a few instances simultaneously. The number of processes per node ranged were 1, 2, 3, 4, 7, or 8. In general, the packets analyzed per process $N_V/N_P$ were $10^{26}$ or $10^{27}$.  For multi-node benchmarks the values highlighted in {\bf bold} in Table~\ref{tab:AnonNetSense-parameters} were used.

\section{Performance Results}

Figure~\ref{fig:BasicFunctions} shows the within node (vertical scaling) and temporal scaling (across time) of the Basic Functions for all the different configurations of hardware listed in Table~\ref{tab:HardwareTable}.    All plots show excellent vertical scaling within a node and temporal scaling over multiple eras of hardware.  These Basic Function performance results set the foundation for running the Anonymized Network Sensing challenge.

Figure~\ref{fig:AnonNetSensing-scaling} (left) shows the within node (vertical scaling) and temporal scaling (across time) of the Anonymized Network Sensing challenge for all the different configurations of hardware listed in Table~\ref{tab:HardwareTable}.  Figure~\ref{fig:AnonNetSensing-scaling} (right) shows the across node (horizontal) and temporal scaling Anonymized Network Sensing challenge, which scales linearly with the number of nodes. These packet per second results are also shown in terms of bits per second using the standard approximation of 1 packet $\approx$ 10,000 bits.  All plots show excellent vertical scaling within a node, horizontal scaling across nodes, and temporal scaling over multiple eras of hardware.

Finally, it is worth mentioning that a popular way to report performance in accelerator hardware is via SpFlops (Sparse Flops), which uses dense operation counts when performing sparse operations.  In some sense, this gives credit for work that was not done.  However, it has also become a popular way to highlight algorithmic improvements in a way that is more broadly appreciated.  The Graph Challenge benchmarks have faced this question for all their benchmarks and have chosen to report the more conservative edge based operation counts.  These are the values listed in Figures~\ref{fig:BasicFunctions} and \ref{fig:AnonNetSensing-scaling}.  For comparison, the peak SpFlops numbers are provided in Table~\ref{tab:SpFlops} and highlights the well-known dramatic performance gains from using sparse matrix operations on sparse data.

\begin{table}
\caption{GigaSpFlops}
\vspace{-0.25cm}
Sparse Flops (SpFlops) equivalent of the best node performance for each benchmark.  All units are in Giga.  Largest values are in the hundreds of ExaSpFlops. Highlights the well-known dramatic performance gains of sparse matrix operations on sparse data.
\begin{center}
\includegraphics[width=\columnwidth]{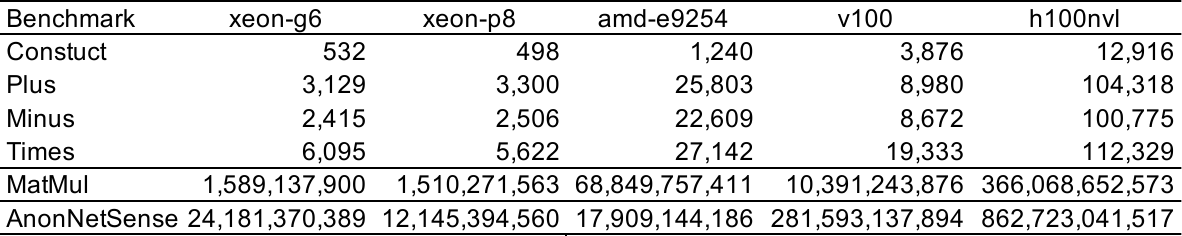}
\end{center}
\label{tab:SpFlops}
\end{table}

\section{Conclusion}

The need to analyze networks with the highest regard for privacy is essential to ensure their proper function and is becoming important as this critical infrastructure expands. The ability to handle diverse data is key requirement as  the network layers use different representations of sources and destinations that can be any combination of physical, logical, or persona/agentic endpoints.  Associative (token) arrays naturally encompass diverse data while still providing scalable mathematical capability. The large scale of modern networks suggest accelerating these libraries with GPUs can be beneficial. The MIT/IEEE/Amazon Anonymized Network Sensing Graph Challenge provides a venue for highlighting the applicability of accelerated associative (token) arrays for these types of problems.  This work benchmarks a prototype Matlab D4M GPU accelerated implementation of the Anonymized Network Sensing challenge across a wide range of CPU and GPU hardware.  Scalable performance is demonstrated within and across CPU cores, CPU nodes, and GPU nodes.  Horizontal scaling across multiple nodes was linear. Running on hundreds of GPU nodes simultaneously achieved a sustained processing rate sufficient to potentially analyze a 10 Petabit/s network.

\section*{Acknowledgement}

The authors wish to acknowledge the following individuals for their contributions and support:
Guillermo Morales, 
S. Atkins, B. Bond, M. Cafarella, B. Cashman, K  Claffy, B. Cook, C. Demchak, A. Edelman, P. Fisher, J. Gottschalk,  T. Hardjono, C. Hill, C. Leiserson, S. Madden, K. Malvey, C. Milner, S. Mohindra, H. Perry, S. Pisharody, C. Prothmann,  S. Rejto, S. Ruppel, D. Rus, P. Saunders, M. Sherman, S. Somin, S. Van Broekhoven, S. Weed.




\bibliographystyle{ieeetr}

\bibliography{AcceleratedTokenArrays}
%

\end{document}